\documentclass[10pt,twocolumn,letterpaper]{article}

\usepackage{wacv}
\usepackage{times}
\usepackage{epsfig}
\usepackage{graphicx}
\usepackage{amsmath}
\usepackage{amssymb}
\usepackage{booktabs}
\usepackage{multirow} 

\def\wacvPaperID{462} 

\wacvalgorithmstrack   

\wacvfinalcopy 

\ifwacvfinal
\usepackage[breaklinks=true,bookmarks=false]{hyperref}
\else
\usepackage[pagebackref=true,breaklinks=true,colorlinks,bookmarks=false]{hyperref}
\fi

\begin{document}

\title{Multimodal Multi-Head Convolutional Attention with Various Kernel Sizes for Medical Image Super-Resolution\vspace*{-0.3cm}}

\author{Mariana-Iuliana Georgescu$^{1}$, Radu Tudor Ionescu$^{1}$, Andreea-Iuliana Miron$^{2,3}$, Olivian Savencu$^{2,3}$,\\ Nicolae-C\u{a}t\u{a}lin Ristea$^{1,4}$, Nicolae Verga$^{2,3}$, Fahad Shahbaz Khan$^{5,6}$\\
$^1$University of Bucharest, Romania, $^2$``Carol Davila'' University of Medicine and Pharmacy, Romania,\\
$^3$Col\c{t}ea Hospital, Romania, $^4$University Politehnica of Bucharest, Romania,\\
$^5$MBZ University of Artificial Intelligence, UAE, $^6$Link\"{o}ping University, Sweden
\vspace*{-0.3cm}
}

\maketitle
\thispagestyle{empty}

\begin{abstract}
\vspace{-0.2cm}
Super-resolving medical images can help physicians in providing more accurate diagnostics. In many situations, computed tomography (CT) or magnetic resonance imaging (MRI) techniques capture several scans (modes) during a single investigation, which can jointly be used (in a multimodal fashion) to further boost the quality of super-resolution results. To this end, we propose a novel multimodal multi-head convolutional attention module to super-resolve CT and MRI scans. Our attention module uses the convolution operation to perform joint spatial-channel attention on multiple concatenated input tensors, where the kernel (receptive field) size controls the reduction rate of the spatial attention, and the number of convolutional filters controls the reduction rate of the channel attention, respectively. We introduce multiple attention heads, each head having a distinct receptive field size corresponding to a particular reduction rate for the spatial attention. We integrate our multimodal multi-head convolutional attention (MMHCA) into two deep neural architectures for super-resolution and conduct experiments on three data sets. Our empirical results show the superiority of our attention module over the state-of-the-art attention mechanisms used in super-resolution. Moreover, we conduct an ablation study to assess the impact of the components involved in our attention module, \eg the number of inputs or the number of heads. Our code is freely available at \url{https://github.com/lilygeorgescu/MHCA}.
\vspace{-0.2cm}
\end{abstract}

\setlength{\abovedisplayskip}{3.0pt}
\setlength{\belowdisplayskip}{3.0pt}

\section{Introduction}
\vspace{-0.1cm}

Magnetic Resonance Imaging (MRI) and Computer Tomography (CT) scanners are non-invasive investigation tools that produce cross-sectional images of various organs or body parts. In common medical practice, the resulting scans are used to diagnose and treat various lesions, ranging from malignant tumors to hemorrhages. Moreover, lesion detection and segmentation from CT and MRI scans are central problems studied in medical imaging, being addressed via automatic techniques \cite{Burduja-S-2020,Shakeel-M-2019,Wang-ACMMM-2017}. However, one voxel in typical MRI or CT scans corresponds to a cubic millimeter of tissue at best, which translates into a rather low resolution, preventing precise diagnosis and treatment. Indeed, according to Georgescu et al.~\cite{Georgescu-ACCESS-2020}, physicians recognize the necessity of increasing the resolution of MRI and CT scans to improve the accuracy of diagnosis and treatment. Furthermore, a recent study \cite{Sert-MH-2019} shows that super-resolution can also aid deep learning models to increase segmentation performance. Due to the aforementioned benefits, we consider that medical image super-resolution (SR) is a very important task for medicine nowadays.

A common medical practice is to take multiple scans with various contrasts (modes) during a single investigation, providing richer information to physicians, who get a more clear picture of the patients. A series of previous works \cite{Feng-MICCAI-2021b,Lyu-TMI-2020,Zeng-CBM-2018,Zheng-BMCMI-2017,Zheng-ACCESS-2018} showed the benefits of using multi-contrast (multimodal) scans to improve super-resolution results. While previous works \cite{Feng-MICCAI-2021b,Lyu-TMI-2020,Zeng-CBM-2018,Zheng-BMCMI-2017,Zheng-ACCESS-2018} combined a low-resolution (LR) scan with a high-resolution (HR) scan of distinct contrasts, to the best of our knowledge, we are the first to study super-resolution with multiple low resolution scans as input. Our approach is applicable to a broader set of CT/MRI scanners, as it does not require the availability of an HR input from another modality (this is rarely available in daily medical practice).

\begin{figure*}[ht]
  \centering
  \includegraphics[width=\linewidth]{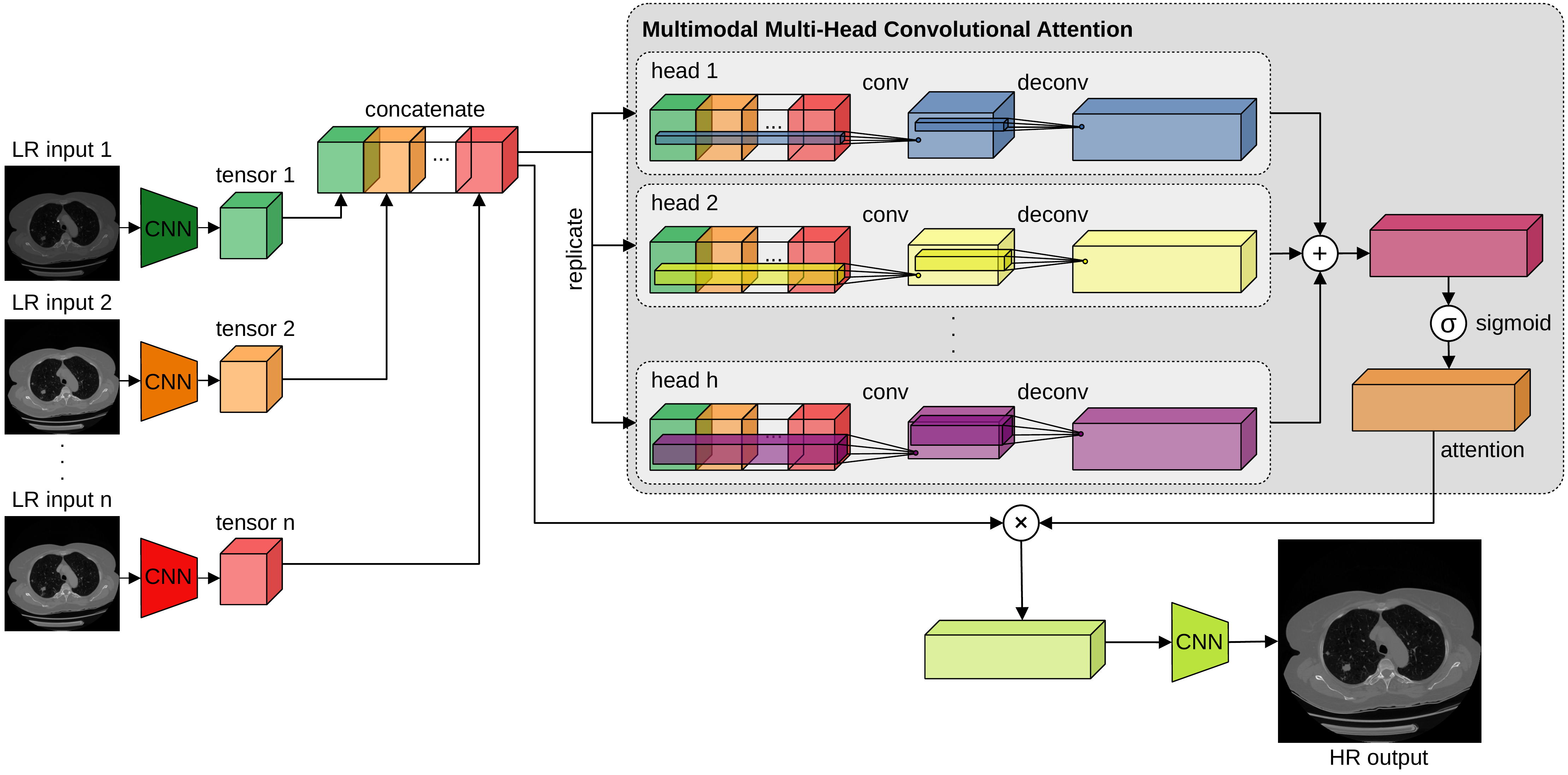}
  \caption{Our multimodal multi-head convolutional attention module (MMHCA) with $h$ heads, integrated into some neural architecture for super-resolution. Input low-resolution (LR) images of distinct contrasts are processed by independent branches and the resulting tensors are concatenated. The concatenated tensor is provided as input to every attention head. Each attention head applies conv and deconv operations, using a kernel size that is unique to the respective head. The resulting tensors are summed up and passed through a sigmoid layer. Finally, the attention tensor is multiplied (element-wise) with the concatenated tensor, and the result is further processed by the network to obtain the high-resolution (HR) image. Best viewed in color.}\label{fig_pipe}
  \vspace{-0.3cm}
\end{figure*}

To approach multi-contrast medical image super-resolution, we propose a novel multimodal multi-head convolutional attention (MMHCA) mechanism that performs joint spatial and channel attention within each head, by stacking a convolutional (conv) layer and a deconvolutional (deconv) layer, as illustrated in Figure~\ref{fig_pipe}. Tensors from different contrast neural branches are concatenated along the channel dimension and given as input to our attention module. The convolutional layer reduces the input tensor both spatially and channel-wise. The channel reduction rate is controlled by adjusting the number of convolutional filters, while the spatial reduction rate is controlled by adjusting the size of the kernel (receptive field). The deconv layer brings the output of the convolutional layer back to its original size. We introduce multiple attention heads, each head having a distinct kernel size corresponding to a particular reduction rate for the spatial attention. The tensors from all attention heads are summed up and passed through a sigmoid layer, obtaining the final attention. The attention tensor is multiplied with the input tensor, enabling the neural network to focus on the most interesting regions from each input image. As other attention modules \cite{Hu-CVPR-2018,Niu-ECCV-2020,Woo-ECCV-2018}, MMHCA relies on the bottleneck principle to force the model in keeping the information that merits attention.

We integrate our multi-input multi-head convolutional attention into two deep neural architectures for super-resolution \cite{Georgescu-ACCESS-2020,Lim-CVPRW-2017} and conduct experiments on three data sets: IXI, NAMIC Multimodality, and Coltea-Lung-CT-100W. These data sets contain multi-contrast investigations which allow us to evaluate our multimodal framework. Our results show that MMHCA brings significant performance gains for both neural networks on all three data sets. Moreover, our framework outperforms recently introduced attention modules \cite{Feng-MICCAI-2021b,Niu-ECCV-2020,Woo-ECCV-2018}, as well as state-of-the-art methods~\cite{Dong-TPAMI-2016,Feng-MICCAI-2021a,Georgescu-ACCESS-2020,Hui-CVPR-2018,Kim-CVPR-2016,Lim-CVPRW-2017,Shi-JBHI-2019,You-TMI-2019,Zeng-CBM-2018,Zhang-CVPR-2021,Zhang-CVPR-2018,Zhao-TIP-2019}. Aside from evaluating methods via automatic measures, \ie the peak signal-to-noise ratio (PSNR) and the structural similarity index measure (SSIM), we conduct a subjective evaluation study, asking three physicians to compare the super-resolution results of a state-of-the-art model, before and after adding MMHCA, without disclosing the method producing each image. The least number of votes assigned by a human annotator to MMHCA is $75\%$, suggesting that its performance gains are indeed significant. In addition, we present ablation results indicating that each component involved in our attention module is important.

In summary, our contribution is threefold:
\begin{itemize}
    \item \vspace{-0.2cm} We are the first to perform medical image super-resolution using a multimodal low-resolution input.
    \item \vspace{-0.2cm} We propose a novel multimodal multi-head convolutional attention mechanism for multi-contrast medical image SR.
    \item \vspace{-0.2cm} We present empirical evidence showing that our attention module brings significant performance gains on three multi-contrast data sets.
\end{itemize}

\section{Related Work}
\vspace{-0.1cm}

\subsection{Image Super-Resolution}
\vspace{-0.1cm}

Most of the recent works~\cite{Du-AS-2019,Feng-MICCAI-2021b,Georgescu-ACCESS-2020,Gu-MTA-2020,Hatvani-TRPMS-2018,Kawulok-GRSL-2020,Kim-CVPR-2016,Lyu-TMI-2020,Niu-ECCV-2020,Shi-CVPR-2016,Yu-ICIP-2017,Zeng-CBM-2018,Zhang-CVPR-2021,Zhang-CVPR-2018,Zhao-TIP-2019,Zheng-BMCMI-2017,Zheng-ACCESS-2018} addressing the super-resolution task use deep learning methods in order to increase the resolution of images. One of the early studies employing deep convolution neural networks (CNNs) for super-resolution is the work of Kim et al.~\cite{Kim-CVPR-2016}, which introduces the Very Deep Super-Resolution (VDSR) network. The VDSR model consists of $20$ convolutional layers and takes as input the interpolated low-resolution (ILR) image. Different from the work of Kim et al.~\cite{Kim-CVPR-2016} which relies on ILR images, Shi et al.~\cite{Shi-CVPR-2016} proposed the efficient sub-pixel convolutional (ESPC) layer, which learns to upscale low-resolution feature maps into the high-resolution output. By eliminating the reliance on ILR images (which have the same height and width as HR images), ESPC is capable of decreasing the running time by a significant margin. After the introduction of the ESPC layer, many researchers~\cite{Du-AS-2019,Feng-MICCAI-2021b,Georgescu-ACCESS-2020,Hatvani-TRPMS-2018,Niu-ECCV-2020,Yu-ICIP-2017,Zhao-TIP-2019} adopted this approach in their super-resolution (SR) models.

Super-resolution methods have also shown their benefits in medical imaging. Medical image super-resolution works can be grouped into two categories, where one category is focused on increasing the resolution of individual CT or MRI slices (2D images)~\cite{Du-NC-2019,Du-AS-2019,Feng-MICCAI-2021b,Feng-MICCAI-2021a,Hatvani-TRPMS-2018,Li-Access-2019,Mahapatra-CMIG-2019,Sert-MH-2019,Shi-JBHI-2019,You-TMI-2019,Yu-ICIP-2017,Zhang-CVPR-2021,Zhao-TIP-2019}, while the other is focused on increasing the resolution of entire 3D scans (volumes)~\cite{Chen-MICCAI-2018,Du-BIBM-2018,Georgescu-ACCESS-2020,Huang-CVPR-2017,Oktay-MICCAI-2016}. Similar to~\cite{Du-NC-2019,Du-AS-2019,Feng-MICCAI-2021b,Feng-MICCAI-2021a,Hatvani-TRPMS-2018,Li-Access-2019,Mahapatra-CMIG-2019,Sert-MH-2019,Shi-JBHI-2019,You-TMI-2019,Yu-ICIP-2017,Zhang-CVPR-2021,Zhao-TIP-2019}, in this work, we are focusing on increasing the resolution of CT and MRI slices. Gu et al.~\cite{Gu-MTA-2020} proposed the MedSRGAN model in order to upsample the resolution of 2D medical images using Generative Adversarial Networks (GANs)~\cite{Goodfellow-NIPS-2014}. MedSRGAN employs a residual map attention network in the generator to extract useful information from different channels. Gu et al.~\cite{Gu-MTA-2020} also used a multi-task loss function comprised of several losses (content loss, adversarial loss and adversarial feature loss) to train the MedSRGAN model. Georgescu et al.~\cite{Georgescu-ACCESS-2020} proposed a method to increase the resolution of both 2D and 3D medical images. To super-resolve 3D images, Georgescu et al.~\cite{Georgescu-ACCESS-2020} used two CNNs in a sequential manner, the first CNN increasing the resolution on two axes (height and width) and the second CNN increasing the resolution on the third axis (depth). Their approach can be used to extend any method from 2D SR to 3D SR, including our own.

Most of the related works employ single-contrast super-resolution (SCSR)~\cite{Du-NC-2019,Du-AS-2019,Georgescu-ACCESS-2020,Hatvani-TRPMS-2018,Li-Access-2019,Mahapatra-CMIG-2019,Sert-MH-2019,Shi-JBHI-2019,You-TMI-2019,Yu-ICIP-2017,Zhang-CVPR-2021,Zhao-TIP-2019}, meaning that they utilize a single-contrast image as input for the upsampling network. There are also some works that approach multi-contrast super-resolution (MCSR)~\cite{Feng-MICCAI-2021b,Lyu-TMI-2020,Zeng-CBM-2018,Zheng-BMCMI-2017,Zheng-ACCESS-2018} using an HR image from another modality, \eg a T1-weighted\footnote{\url{https://radiopaedia.org/articles/mri-sequences-overview}} slice, to increase the resolution of the targeted LR modality, \eg a T2-weighted slice. Zeng et al.~\cite{Zeng-CBM-2018} proposed a model consisting of two sub-networks to simultaneously perform SCSR and MCSR. The first sub-network performs SCSR, upsampling the target modality, while the second sub-network uses the output of the first sub-network and an HR image from a different modality to further refine the target modality. Feng et al.~\cite{Feng-MICCAI-2021b} proposed a multi-stage integration network (MINet) for MCSR. The MINet model uses two input images that are processed in parallel by two independent networks, and their features are fused at each layer to obtain multi-stage feature representations. Similar to Zeng et al.~\cite{Zeng-CBM-2018}, Feng et al.~\cite{Feng-MICCAI-2021b} integrated a second HR modality to increase the performance of their model. Different from previous works~\cite{Feng-MICCAI-2021b,Lyu-TMI-2020,Zeng-CBM-2018,Zheng-BMCMI-2017,Zheng-ACCESS-2018}, we do not employ any HR image to guide our model. Instead, we only rely on the low-resolution images pertaining to different modalities. To the best of our knowledge, we are the first to propose an MCSR method based solely on LR medical images as input.

\subsection{Attention Mechanism}
\vspace{-0.1cm}

The attention mechanism is a very hot topic in the computer vision community, having broad applications ranging from mainstream computer vision tasks, such as image classification~\cite{Woo-ECCV-2018}, to more specific tasks, such as natural image SR~\cite{Niu-ECCV-2020} and medical image SR~\cite{Feng-MICCAI-2021b, Zhang-CVPR-2021}. Attention mechanisms are integrated into neural networks to direct the attention of the model to the area with relevant information. Hu et al.~\cite{Hu-CVPR-2018} proposed the squeeze-and-excitation (SE) block to recalibrate the channel responses, thus performing channel-wise attention. In order to further increase the power of the attention mechanism, Woo et al.~\cite{Woo-ECCV-2018} proposed the Convolutional Block Attention Module (CBAM), which infers attention maps for two separate dimensions, spatial and channel, in a sequential manner. Niu et al.~\cite{Niu-ECCV-2020} proposed the channel-spatial attention module (CSAM) to boost the performance of natural image SR. CSAM is composed of 3D conv layers, being able to learn the channel and spatial interdependencies of the features. In a similar fashion to Niu et al.~\cite{Niu-ECCV-2020}, Feng et al.~\cite{Feng-MICCAI-2021b} employed 3D conv layers to generate attention maps that capture both channel and spatial information. Unlike previous works~\cite{Feng-MICCAI-2021b,Niu-ECCV-2020,Woo-ECCV-2018}, our attention module is based on 2D convolution and deconvolution operations with multiple kernel sizes to perform joint multi-head spatial-channel attention.

The self-attention mechanism \cite{Vaswani-NIPS-2017} triggered the development of models solely based on attention, such as vision transformers \cite{Carion-ECCV-2020,Chen-arXiv-2021,Dosovitskiy-ICLR-2020,Gao-MICCAI-2021,Khan-ACS-2021,Parmar-ICML-2018,Ristea-A-2021,Touvron-ICML-2021,Wu-ICCV-2021,Zheng-BMVC-2021,Zhu-ICLR-2020}, which have been adopted at an astonishing rate by the computer vision and medical imaging communities, likely due to the impressive results across a wide range of problems, from object recognition \cite{Dosovitskiy-ICLR-2020,Touvron-ICML-2021,Wu-ICCV-2021} and object detection \cite{Carion-ECCV-2020,Zheng-BMVC-2021,Zhu-ICLR-2020} to medical image segmentation \cite{Chen-arXiv-2021,Gao-MICCAI-2021,Hatamizadeh-WACV-2022} and medical image generation \cite{Ristea-A-2021}. Although transformers \cite{Shamshad-A-2022} have been applied to several mainstream medical imaging tasks, the number of transformer-based methods applied to medical image super-resolution is relatively small \cite{Feng-A-2021,Feng-MICCAI-2021a}. Different from methods relying solely on attention-based architectures \cite{Feng-A-2021,Feng-MICCAI-2021a}, we propose a novel and flexible attention module that can be integrated into various architectures. To support this claim, we integrate MMHCA into two state-of-the-art architectures \cite{Georgescu-ACCESS-2020,Lim-CVPRW-2017}, providing empirical evidence showing that our attention module can significantly boost the performance of both models.

\section{Method}
\vspace{-0.1cm}

Given a multi-contrast input formed of $n$ low-resolution input images of $p \times p$ pixels, denoted as $LR_1$, $LR_2$, ..., $LR_n$, our goal is to obtain an HR image of $r \times r$ pixels, denoted as $HR_k$, where $r > p$, for the target modality $k \in \{1,2,...,n\}$. In common practice, $r$ is typically chosen to be equal to $2p$ or $4p$, corresponding to super-resolution factors of $2\times$ or $4\times$, respectively. In our experiments, we consider these commonly-used SR factors.

We propose a spatial-channel attention module to combine the information contained by the multi-contrast LR images. As illustrated in Figure~\ref{fig_pipe}, our multimodal multi-head convolutional attention (MMHCA) mechanism can be introduced at any layer of any neural architecture, thus being generic and flexible. We underline that if the baseline architecture is designed for a single-contrast input, we can easily extend the architecture to a multi-contrast input of $n$ contrasts by replicating the neural branch that comes before our module, for a number of $n$ times. Let $f_i$ be the neural branch that processes the input $LR_i$. We first obtain the encoding tensor $T_i$ for each input $LR_i$, as follows:
\begin{equation} 
T_i = f_i\left(\theta_{f_i}, LR_i\right),
\end{equation}
where $\theta_{f_i}$ are the weights of the neural branch $f_i$.

The next step in our approach is to concatenate the encoding tensors $T_i$ of all modalities along the channel axis, obtaining the tensor $T_{\circ}$, as follows:
\begin{equation} 
T_{\circ} = concat\left(T_1, T_2, ..., T_n\right),
\end{equation}
where $concat$ represents the concatenation operation. If a multimodal input is not available, our multi-head convolutional attention (MHCA) module can still be applied by considering $T_{\circ} = T_1$, where $T_1$ is the encoding tensor of the single-contrast input. We use this ablated configuration in our experiments to show the benefits of using multiple modalities as opposed to a single modality.

Next, we apply the multi-head convolutional attention. Our attention module is composed of multiple heads, which are applied on the concatenation of the encoding tensors, denoted as $T_{\circ}$. Each head $h_j$, $j \in \{1, 2, ..., h\}$, where $h$ is the number of heads, is composed of a conv layer followed by a deconv layer, performing both spatial and channel attention. The conv layer jointly reduces the spatial and channel dimensions of the input, while the deconv layer is configured to revert the dimensional reduction performed by the conv layer, thus restoring the size of the output tensor to the size of the input tensor $T_{\circ}$.

For the $j$-th convolutional attention head $h_j$, we set the conv kernel size $k_j$ to $2 \cdot (j - 1) + 1$. Hence, the first head is formed of kernels having a receptive field of $1\times1$, the second head is formed of kernels having a receptive field of $3\times3$, and so on. We note that each kernel size corresponds to a different reduction rate of the spatial attention. For each head $h_j$, we set the number of conv filters to $\frac{c}{r}$, where $c$ is the number of input channels and $r$ is the reduction rate of the channel attention. Then, we apply a deconv layer with $c$ filters and the kernel size set to $k_j \times k_j$ to increase the spatial size of the activation maps. We use a stride of $1$ and a padding equal to $0$ for both conv and deconv layers. In summary, the output $H_j$ of the attention head $h_j$ is computed as follows:
\begin{equation} 
H_j = h_j\left(\theta_{c_j}, \theta_{d_j}, T_{\circ} \right) = max \left(0, T_{\circ} \ast \theta_{c_j} \right) \circledast  \theta_{d_j}, 
\end{equation}
where $\theta_{c_j}$ are the learnable weights of the conv layer comprising $\frac{c}{r}$ filters with a kernel size of $k_j \times k_j$, $\theta_{d_j}$ are the parameters of the deconv layer comprising $c$ filters with a kernel size of $k_j \times k_j$, $max(0,\cdot)$ is the ReLU activation function, $\ast$ is the convolution operation and $\circledast$ is the deconvolution operation. We hereby note that the tensors $H_1$, $H_2$, ..., $H_h$ are of the same size.

Next, we sum all the tensors $H_j$ produced by the attention heads, obtaining $H_+$, as follows:
\begin{equation} 
H_+ = \sum_{j=1}^h H_j.   
\end{equation}
We then pass the resulting tensor through the sigmoid function in order to obtain the final attention tensor denoted as $A$, as follows:
\begin{equation} 
A = \sigma \left(H_+\right).
\end{equation}

To apply the learned attention to the encoding tensors, the attention tensor $A$ is multiplied with the tensor $T_{\circ}$, obtaining the tensor $T_*$, as follows:
\begin{equation} 
T_* = T_{\circ} \otimes A,
\end{equation}
where $\otimes$ denotes the element-wise multiplication operation.

Let $g$ denote the neural branch that takes $T_*$ as input and produces the high-resolution output. The last processing required to obtain the final output of the neural architecture is expressed as follows:
\begin{equation} 
HR_k = g \left(\theta_g, T_*\right),
\end{equation}
where $\theta_g$ are the learnable weights of $g$.

\section{Experiments}
\vspace{-0.1cm}

\subsection{Data Sets}
\vspace{-0.1cm}

\noindent
\textbf{IXI.} The IXI\footnote{\url{http://brain-development.org/ixi-dataset/}} data set is the largest benchmark considered in our evaluation. We use the same version of the IXI data set as \cite{Zhang-CVPR-2021, Zhao-TIP-2019}. The data set contains 3D multimodal MRI scans, where each MRI has three modalities, namely T1-weighted (T1w), T2-weighted (T2w) and Proton Density (PD). The data set is split into $500$ multimodal MRI scans for training, $6$ multimodal scans for validation and the remaining $70$ multimodal scans for testing. Each MRI scan has $96$ slices with the resolution of $240\times240$ pixels.

\noindent
\textbf{NAMIC Brain Multimodality.} The National Alliance for Medical Image Computing (NAMIC) Brain Multimodality\footnote{\url{https://www.na-mic.org/wiki/Downloads}} data set  is formed of $20$ 3D MRI scans. Each 3D image is formed of $176$ slices with the resolution of $256 \times 256$ pixels. The data set contains two modalities, namely T1w and T2w. Following~\cite{Georgescu-ACCESS-2020, Pham-CMIG-2019}, we randomly split the data set into $10$ multimodal MRI scans for training and $10$ multimodal MRI scans for testing. We randomly take $2$ MRI scans from the training set to create a validation set for hyperparameter tuning.

\noindent
\textbf{Coltea-Lung-CT-100W.} The Coltea-Lung-CT-100W data set was recently introduced by Ristea et al.~\cite{Ristea-A-2021}. It contains $100$ triphasic (multimodal) lung CT scans. Each scan has three modalities, namely native, arterial and venous. The entire data set is formed of $12,\!430$ triphasic slices and each slice has $512 \times 512$ pixels. The data set is split into $70$ multimodal CT scans for training, $15$ multimodal scans for validation and $15$ multimodal scans for testing. 

\subsection{Evaluation Metrics}
\vspace{-0.1cm}

As evaluation metrics, we employ the peak signal-to-noise ratio (PSNR) and the structural similarity index measure (SSIM)~\cite{Wang-TIP-2004}. PSNR is the ratio between the maximum possible signal and the power of the noise. It only takes into account the difference between pixels, without quantifying the structural similarity. SSIM~\cite{Wang-TIP-2004} takes the structural similarity into account by combining the contrast, the luminance and the texture of the images. Higher values of PSNR and SSIM indicate better reconstruction. We emphasize that PSNR is represented in the log-scale. Hence, seemingly small gains in terms of PSNR can indicate significant quality improvements.

To further assess the improvement brought by our attention module, we also conduct a subjective evaluation study, asking three physicians (specialized in radiology) from the Col\c{t}ea Hospital to compare the super-resolution results of a state-of-the-art model, before and after adding MMHCA. 

\subsection{Implementation Details}
\vspace{-0.1cm}

We compare our attention module with CSAM \cite{Feng-MICCAI-2021b,Niu-ECCV-2020} and CBAM \cite{Woo-ECCV-2018}. We consider EDSR \cite{Lim-CVPRW-2017} and CNN+ESPC \cite{Georgescu-ACCESS-2020} as underlying models for the attention mechanisms.
To train the EDSR\footnote{\url{https://github.com/sanghyun-son/EDSR-PyTorch}} and CNN+ESPC\footnote{\url{https://github.com/lilygeorgescu/3d-super-res-cnn}} models, we use the official code released by the corresponding authors. For the EDSR \cite{Lim-CVPRW-2017} network, we set $B=16$ and $F=64$. All the other hyperparameters are left unchanged. For CNN+ESPC \cite{Georgescu-ACCESS-2020}, we use the same hyperparameters as suggested by the authors. 

For a fair comparison, we integrate all the attention modules in the same manner into both networks. For the experiments with single-contrast inputs, we integrate the modules (CSAM, CBAM, MHCA) after each ResBlock. For the experiments with multi-contrast inputs, we replicate the sub-network which ends just before the upsampling layer, creating copies of the sub-network (each copy having its own learnable weights). Then, we concatenate the output of each sub-network and introduce the attention modules (MCSAM, MCBAM, MMHCA). The multi-contrast inputs are used without prior alignment.

For MHCA/MMHCA, we tune the number of heads $h$ and the channel reduction ratio $r$ on the validation set of each benchmark. We find that the optimal configuration is based on $h=3$ heads and a channel reduction ratio of $r=0.5$. This configuration uses kernel sizes of $1\times1$ for $h_1$, $3\times3$ for $h_2$, and $5\times5$ for $h_3$, respectively.

When comparing our attention modules (MHCA/MMHCA) with CSAM/MCSAM\footnote{\url{https://github.com/wwlCape/HAN},  \url{https://github.com/chunmeifeng/MINet}} \cite{Feng-MICCAI-2021b,Niu-ECCV-2020} and CBAM/MCBAM\footnote{\url{https://github.com/Jongchan/attention-module}} \cite{Woo-ECCV-2018}, we use the official code released by the respective authors. As for MHCA/MMHCA, we tune the hyperparameters of CBAM/MCBAM on the validation sets, while CSAM/MCSAM have no hyperparameters that would require tuning.
The optimal configuration for CBAM/MCBAM uses a kernel size of $7\times7$ and a reduction ratio of $0.5$.

\begin{figure*}[!th]
  \centering
  \includegraphics[width=0.755\linewidth]{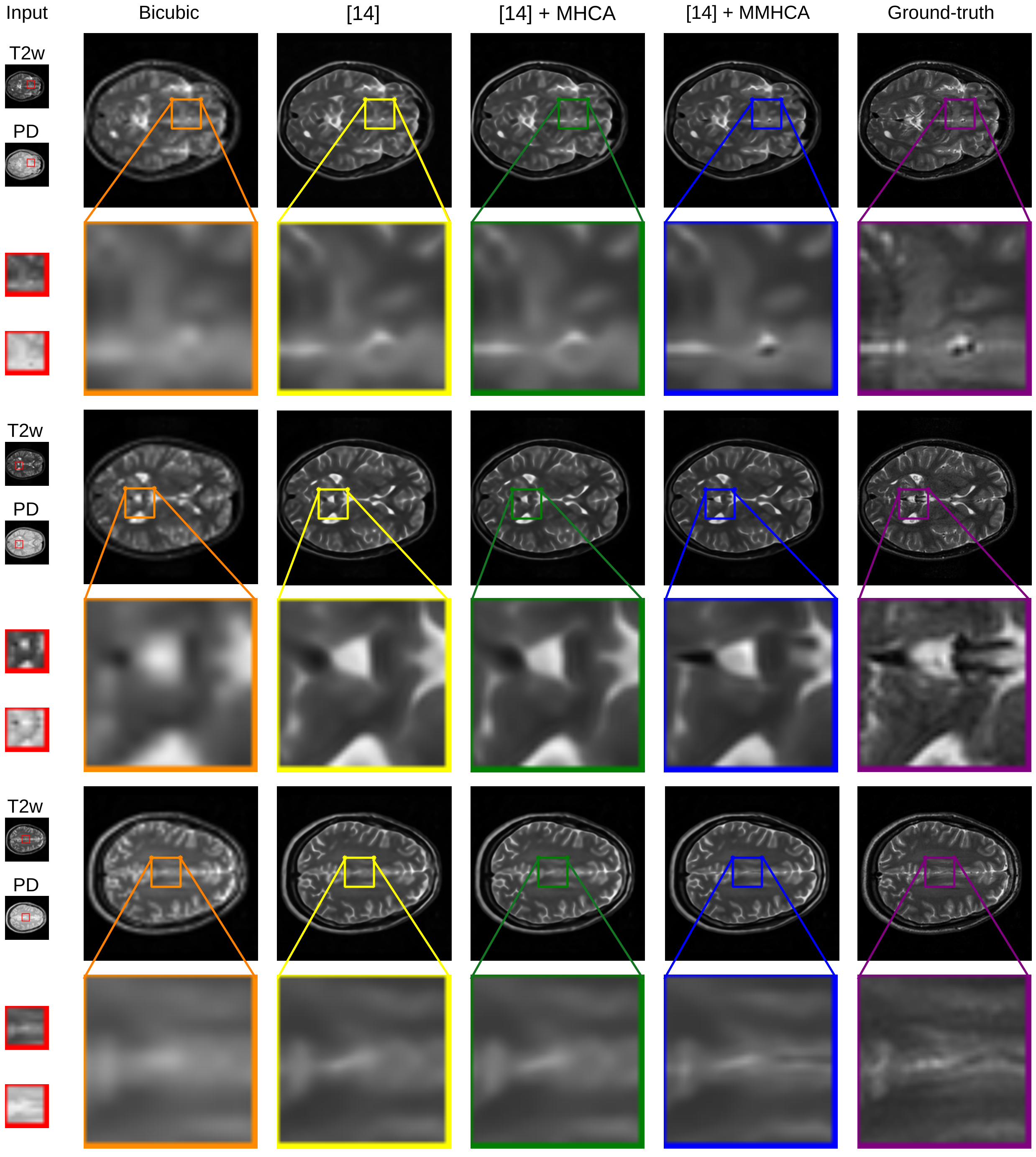}
  \vspace{-0.1cm}
  \caption{Examples of super-resolved MRI slices from the IXI data set, for an upscaling factor of $4\times$. The HR images produced by two baselines (bicubic interpolation and CNN+ESPC \cite{Georgescu-ACCESS-2020}) are compared with the images given by two enhanced versions of CNN+ESPC \cite{Georgescu-ACCESS-2020}, one based on our single-contrast attention module (MHCA), and another based on our multimodal attention module (MMHCA).}\label{fig_2}
  \vspace{-0.3cm}
\end{figure*}

\begin{table}
  \begin{center}
\footnotesize{
  \setlength\tabcolsep{2.5pt}
  
  \begin{tabular}{lcc}
    \toprule
    \multirow{2}{*}{Method} &  $2\times$ &  $4\times$\\
            & PSNR/SSIM & PSNR/SSIM \\
    \midrule
    Bicubic                     &   $33.44 / 0.9589$  & $27.86 / 0.8611$\\
    \hline
    SRCNN \cite{Dong-TPAMI-2016}  &   $37.32 / 0.9796$  & $29.69 / 0.9052$\\
    VDSR \cite{Kim-CVPR-2016}     &   $38.65 / 0.9836$  & $30.79 / 0.9240$\\
    IDN \cite{Hui-CVPR-2018}      &   $39.09 / 0.9846$  & $31.37 / 0.9312$\\
    RDN \cite{Zhang-CVPR-2018}    &   $38.75 / 0.9838$  & $31.45 / 0.9324$\\
    FSCWRN \cite{Shi-JBHI-2019}   &   $39.44 / 0.9855$  & $31.71 / 0.9359$\\
    CSN \cite{Zhao-TIP-2019}      &   $39.71 / 0.9863$  & $32.05 / 0.9413$\\
    T$^2$Net \cite{Feng-MICCAI-2021a} &   $29.38 / 0.8720$  & $28.66 / 0.8500$\\ 
    SERAN \cite{Zhang-CVPR-2021}  &   $40.18 / 0.9872$  & $32.40 / 0.9455$\\ 
    SERAN$^+$ \cite{Zhang-CVPR-2021}  &   ${\color{blue}40.30} / {\color{blue}0.9874}$  & ${\color{blue}32.62} / {\color{red}0.9472}$\\ 
    \hline
    EDSR \cite{Lim-CVPRW-2017}    &   $39.81 / 0.9865$  & $31.83 / 0.9377$\\ 
    EDSR \cite{Lim-CVPRW-2017} + CSAM \cite{Feng-MICCAI-2021b,Niu-ECCV-2020}  &  $39.81 / 0.9865$    & $31.83 / 0.9377$ \\
    EDSR \cite{Lim-CVPRW-2017} + CBAM \cite{Woo-ECCV-2018}           &  $39.82 / 0.9865$    & $31.81 / 0.9374$ \\
    EDSR \cite{Lim-CVPRW-2017} + MHCA (ours)  &  $40.11 / 0.9871$  & $32.15 / 0.9418$\\ 
    \hline
    EDSR \cite{Lim-CVPRW-2017} + MCSAM \cite{Feng-MICCAI-2021b,Niu-ECCV-2020}     &  $40.12 / 0.9871$    & $32.17 / 0.9417$ \\
    EDSR \cite{Lim-CVPRW-2017} + MCBAM \cite{Woo-ECCV-2018}       &  $40.13 / 0.9871$    & $32.18 / 0.9421$ \\ 
    EDSR \cite{Lim-CVPRW-2017} + MMHCA (ours)   &  $40.28 / {\color{blue}0.9874}$  & $32.51 / 0.9452$\\  
    EDSR \cite{Lim-CVPRW-2017} + MMHCA$^+$ (ours)   &  ${\color{red}40.43} / {\color{red}0.9877}$  & ${\color{red}32.70} / {\color{blue}0.9469}$\\  
    \hline
    CNN+ESPC \cite{Georgescu-ACCESS-2020}          &  $38.67 / 0.9837$  & $30.57 / 0.9210$\\
    CNN+ESPC \cite{Georgescu-ACCESS-2020} + CSAM \cite{Feng-MICCAI-2021b,Niu-ECCV-2020}                 &  $38.57 / 0.9835$    & $30.58 / 0.9211$ \\
    CNN+ESPC \cite{Georgescu-ACCESS-2020} + CBAM \cite{Woo-ECCV-2018}                &  $38.67 / 0.9838$    & $30.47 / 0.9192$ \\
    CNN+ESPC \cite{Georgescu-ACCESS-2020} + MHCA (ours)   &  $39.04 / 0.9847$  & $30.76 / 0.9233$\\
    \hline
    CNN+ESPC \cite{Georgescu-ACCESS-2020} + MCSAM \cite{Feng-MICCAI-2021b,Niu-ECCV-2020}    &  $38.98 / 0.9845$    & $30.94 / 0.9265$ \\
    CNN+ESPC \cite{Georgescu-ACCESS-2020} + MCBAM  \cite{Woo-ECCV-2018}   &  $38.91 / 0.9844$    & $30.79 / 0.9238$ \\
    CNN+ESPC \cite{Georgescu-ACCESS-2020} + MMHCA (ours) &   $39.71 / 0.9862$  & $31.52 / 0.9337$\\
    CNN+ESPC \cite{Georgescu-ACCESS-2020} + MMHCA$^+$ (ours) &   $39.76 / 0.9863$  & $31.52 / 0.9337$\\  
  \bottomrule
\end{tabular}
}
\end{center}
\vspace{-0.1cm}
\caption{PSNR and SSIM scores of various state-of-the-art methods \cite{Dong-TPAMI-2016,Feng-MICCAI-2021a,Georgescu-ACCESS-2020,Hui-CVPR-2018,Kim-CVPR-2016,Lim-CVPRW-2017,Shi-JBHI-2019,Zhang-CVPR-2018,Zhang-CVPR-2021,Zhao-TIP-2019} on the IXI data set, for the T2w target modality. For two of the existing methods (EDSR \cite{Lim-CVPRW-2017} and CNN+ESPC \cite{Georgescu-ACCESS-2020}), we evaluate enhanced versions, considering various state-of-the-art attention modules \cite{Feng-MICCAI-2021b,Niu-ECCV-2020,Woo-ECCV-2018}, as well as our own attention module (MMHCA). We consider both single-contrast (CSAM, CBAM, MHCA) and multi-contrast (MCSAM, MCBAM, MMHCA, MMHCA$^+$) versions. The top two scores for each scaling factor ($2\times$ and $4\times$) are highlighted in red and blue, respectively.}
  \label{tab_ixi}
\end{table}

\begin{table} 
   \begin{center}
\footnotesize{
  \setlength\tabcolsep{2.5pt}
  \begin{tabular}{lcc}
    \toprule
    \multirow{2}{*}{Method} &  $2\times$ &  $4\times$\\
            & PSNR/SSIM & PSNR/SSIM \\
    \midrule
    Bicubic                         &  $37.61 / 0.9794$    & $31.35 / 0.9091$ \\
    \hline
    MCSR~\cite{Zeng-CBM-2018} &  $38.32 / 0.9450$    & $30.84 / 0.8110$ \\
    GAN-CIRCLE~\cite{You-TMI-2019}  &  $36.19 / 0.9594 $    & $- $ \\ 
    \hline
    EDSR \cite{Lim-CVPRW-2017} &  $41.46 / 0.9906$    & $34.50 / 0.9558$ \\ 
    EDSR \cite{Lim-CVPRW-2017} + CSAM \cite{Feng-MICCAI-2021b,Niu-ECCV-2020}   &  $41.43 / 0.9905$    & $34.53	/ 0.9560$ \\
    EDSR \cite{Lim-CVPRW-2017} + CBAM \cite{Woo-ECCV-2018}                     &  $38.66 / 0.9840$    & $32.59 / 0.9328$ \\
    EDSR \cite{Lim-CVPRW-2017} + MHCA (ours) &  $41.94 / 0.9914$    & $34.99	/ 0.9601$ \\
    \hline  
    EDSR \cite{Lim-CVPRW-2017} + MCSAM \cite{Feng-MICCAI-2021b,Niu-ECCV-2020} &  $41.52 / 0.9907$    & $34.66	/ 0.9571$ \\
    EDSR \cite{Lim-CVPRW-2017} + MCBAM \cite{Woo-ECCV-2018} &  $41.49 / 0.9906$    & $34.65	/ 0.9569$ \\
    EDSR \cite{Lim-CVPRW-2017} + MMHCA (ours) &  ${\color{blue}42.15} / {\color{blue}0.9917}$    & ${\color{blue}35.35} / {\color{blue}0.9624}$ \\
    EDSR \cite{Lim-CVPRW-2017} + MMHCA$^+$ (ours)   &  ${\color{red}42.32} / {\color{red}0.9919}$  & ${\color{red}35.58} / {\color{red}0.9638}$\\  
    \hline  
    CNN+ESPC \cite{Georgescu-ACCESS-2020}  &  $40.73 / 0.9893$    & $33.92 / 0.9509$ \\
    CNN+ESPC \cite{Georgescu-ACCESS-2020} + CSAM \cite{Feng-MICCAI-2021b,Niu-ECCV-2020}  &  $40.46 / 0.9888$    & $33.86	/ 0.9503$ \\
    CNN+ESPC \cite{Georgescu-ACCESS-2020} + CBAM \cite{Woo-ECCV-2018}   &  $40.54 / 0.9889$    & $33.60 / 0.9492$ \\
    CNN+ESPC \cite{Georgescu-ACCESS-2020} + MHCA (ours) &  $40.81 / 0.9894$    & $33.94 / 0.9511$\\
    \hline   
    CNN+ESPC \cite{Georgescu-ACCESS-2020} + MCSAM \cite{Feng-MICCAI-2021b,Niu-ECCV-2020} &  $40.74 / 0.9893$    & $34.04 / 0.9518$ \\
    CNN+ESPC \cite{Georgescu-ACCESS-2020} + MCBAM  \cite{Woo-ECCV-2018}   &  $40.73 / 0.9892$    & $33.88	/ 0.9516$ \\
    CNN+ESPC \cite{Georgescu-ACCESS-2020} + MMHCA (ours)&  $41.58 / 0.9908$    & $34.64 / 0.9573$ \\
    CNN+ESPC \cite{Georgescu-ACCESS-2020} + MMHCA$^+$ (ours) &   $41.49 / 0.9904$  & $34.70 / 0.9573$\\  
  \bottomrule
 
\end{tabular}
}
  \end{center}
  \vspace{-0.1cm}
      \caption{PSNR and SSIM scores of various state-of-the-art methods \cite{Georgescu-ACCESS-2020,Lim-CVPRW-2017,You-TMI-2019,Zeng-CBM-2018} on the NAMIC data set, for the T2w target modality. For two of the existing methods (EDSR \cite{Lim-CVPRW-2017} and CNN+ESPC \cite{Georgescu-ACCESS-2020}), we evaluate enhanced versions, considering various state-of-the-art attention modules \cite{Feng-MICCAI-2021b,Niu-ECCV-2020,Woo-ECCV-2018}, as well as our own attention module (MMHCA). We consider both single-contrast (CSAM, CBAM, MHCA) and multi-contrast (MCSAM, MCBAM, MMHCA, MMHCA$^+$) versions. The top two scores for each scaling factor ($2\times$ and $4\times$) are highlighted in red and blue, respectively.}  \label{tab_namic}
\end{table}

\begin{table}
   \begin{center}
\footnotesize{
  \setlength\tabcolsep{2.5pt}
  \begin{tabular}{lcc}
    \toprule
    \multirow{2}{*}{Method} &  $2\times$ &  $4\times$\\
            & PSNR/SSIM & PSNR/SSIM \\
    \midrule
    Bicubic                         &  $38.84 / 0.9477$    & $31.47 / 0.8774$ \\
    \hline
    EDSR \cite{Lim-CVPRW-2017}                            &  $45.11 / 0.9621$    & $39.52 / 0.9394$ \\
    
    EDSR \cite{Lim-CVPRW-2017} + CSAM \cite{Feng-MICCAI-2021b,Niu-ECCV-2020} &  $45.12 / 0.9622$    & $39.57 / 0.9395$ \\
    EDSR \cite{Lim-CVPRW-2017} + CBAM \cite{Woo-ECCV-2018}                    &  $41.14 / 0.9538$    & $32.92 / 0.9023$\\ 
    EDSR \cite{Lim-CVPRW-2017} + MHCA (ours)        &  $45.50 / 0.9634$    & ${\color{red}40.23} / {\color{red}0.9416}$ \\
    \hline
    EDSR \cite{Lim-CVPRW-2017} + MCSAM \cite{Feng-MICCAI-2021b,Niu-ECCV-2020}  & $45.16 / 0.9623$    & $39.62 / 0.9397$\\
    EDSR \cite{Lim-CVPRW-2017} + MCBAM \cite{Woo-ECCV-2018}   & $45.12 / 0.9622$    & $39.65 / 0.9397$ \\ 
    EDSR \cite{Lim-CVPRW-2017} + MMHCA (ours) & ${\color{blue}45.58} / {\color{blue}0.9647}$    & $40.06 / 0.9404$ \\
    EDSR \cite{Lim-CVPRW-2017} + MMHCA$^+$ (ours)   &  ${\color{red}45.68} / {\color{red}0.9649}$  & ${\color{blue}40.22} / {\color{blue}0.9409}$\\  
    \hline
    CNN+ESPC \cite{Georgescu-ACCESS-2020}   & $44.47 / 0.9599$    & $38.34 / 0.9338$\\
    CNN+ESPC \cite{Georgescu-ACCESS-2020} + CSAM \cite{Feng-MICCAI-2021b,Niu-ECCV-2020} & $44.45 / 0.9599$    & $38.36 / 0.9337$ \\
    CNN+ESPC \cite{Georgescu-ACCESS-2020} + CBAM \cite{Woo-ECCV-2018} & $44.44 / 0.9601$    & $37.23 / 0.9308$ \\ 
    CNN+ESPC \cite{Georgescu-ACCESS-2020} + MHCA (ours) & $44.72 / 0.9608$    & $38.62 / 0.9348$ \\
    \hline
    CNN+ESPC \cite{Georgescu-ACCESS-2020} + MCSAM \cite{Feng-MICCAI-2021b,Niu-ECCV-2020} & $44.51 / 0.9599$    & $38.38 / 0.9338$\\
    CNN+ESPC \cite{Georgescu-ACCESS-2020} + MCBAM \cite{Woo-ECCV-2018} & $44.43 / 0.9600$    & $38.19 / 0.9318$ \\ 
    CNN+ESPC \cite{Georgescu-ACCESS-2020} + MMHCA (ours) & $45.05 / 0.9621$    & $38.96 / 0.9365$ \\ 
    CNN+ESPC \cite{Georgescu-ACCESS-2020} + MMHCA$^+$ (ours) &   $45.14 / 0.9623$  & $39.10 / 0.9370$\\  
  \bottomrule
\end{tabular}
}
\end{center}
\vspace{-0.1cm}
 \caption{PSNR and SSIM scores of two state-of-the-art methods \cite{Georgescu-ACCESS-2020,Lim-CVPRW-2017} on the Coltea-Lung-CT-100W data set, for the native target modality. We evaluate enhanced versions of the existing methods, considering various state-of-the-art attention modules \cite{Feng-MICCAI-2021b,Niu-ECCV-2020,Woo-ECCV-2018}, as well as our own attention module (MMHCA). We consider both single-contrast (CSAM, CBAM, MHCA) and multi-contrast (MCSAM, MCBAM, MMHCA, MMHCA$^+$) versions. The top two scores for each scaling factor ($2\times$ and $4\times$) are highlighted in red and blue, respectively.}
  \label{tab_ch}
\end{table}


\begin{table}
       \begin{center}
\footnotesize{
  \begin{tabular}{lcc}
    \toprule
    Method & EDSR \cite{Lim-CVPRW-2017} & EDSR \cite{Lim-CVPRW-2017} + MMHCA (ours) \\
    \midrule
    Doctor $\#1$ &  $11$ &   $89$ \\
    Doctor $\#2$ &  $25$ &   $75$ \\
    Doctor $\#3$ &  $20$ &   $80$ \\
    \hline
    Average (\%) & $18.6\%$ &  $\mathbf{81.3}\%$ \\
  \bottomrule
\end{tabular} 
}
\end{center}
\vspace{-0.1cm}
 \caption{Subjective human evaluation results based on 100 randomly selected cases from the IXI test set for EDSR \cite{Lim-CVPRW-2017}, with and without MMHCA, considering an upscaling factor of $4\times$. The reported numbers represent votes awarded by three physicians (with expertise in radiology) for each model.}
 \label{tab_human}
\end{table}

\begin{table}
     \begin{center}
\footnotesize{
  \setlength\tabcolsep{2.5pt}
  
  \begin{tabular}{lc}
    \toprule
    Method & PSNR/SSIM \\
    \midrule
    EDSR \cite{Lim-CVPRW-2017}                        & $41.46 / 0.9906$    \\
    EDSR \cite{Lim-CVPRW-2017} + MHCA (1 head, r = $0.5$)                   & $41.40 / 0.9904$    \\
    EDSR \cite{Lim-CVPRW-2017} + MHCA (2 heads, r = $0.5$)                   & $41.79 / 0.9911$    \\
    EDSR \cite{Lim-CVPRW-2017} + MHCA (3 heads, r = $0.5$)                   & $41.94 / 0.9914$    \\
    EDSR \cite{Lim-CVPRW-2017} + MHCA (4 heads, r = $0.5$)                   & $41.90 / 0.9913$    \\
    \hline
    EDSR \cite{Lim-CVPRW-2017} + MHCA (3 heads, no deconv, r = $0.5$)                   & $41.90 / 0.9913$    \\ 
    \hline
    EDSR \cite{Lim-CVPRW-2017} + MHCA (4 heads, $1\!\times\!1$ kernels, r = $2$)                   & $41.04 / 0.9898$    \\ 
    EDSR \cite{Lim-CVPRW-2017} + MHCA (4 heads, r = $2$)                   & $41.89 / 0.9913$    \\ 
    \hline
    EDSR \cite{Lim-CVPRW-2017} + multimodal input            & $41.50 /	0.9907$    \\
    EDSR \cite{Lim-CVPRW-2017} + MMHCA (1 head, r = $0.5$)          & $41.76 / 0.9911$    \\
    EDSR \cite{Lim-CVPRW-2017} + MMHCA (2 heads, r = $0.5$)        & $42.05 / 0.9915$    \\
    EDSR \cite{Lim-CVPRW-2017} + MMHCA (3 heads, r = $0.5$)        & $42.15 / 0.9917$    \\
    EDSR \cite{Lim-CVPRW-2017} + MMHCA (4 heads, r = $0.5$)        & $42.09 / 0.9916$    \\
    \hline
    EDSR \cite{Lim-CVPRW-2017} + MMHCA (3 heads, no deconv, r = $0.5$)                   & $41.87 / 0.9912$    \\
    \hline
    EDSR \cite{Lim-CVPRW-2017} + MMHCA (4 heads, $1\!\times\!1$ kernels, r = $2$)                   & $41.73 / 0.9910$    \\ 
    EDSR \cite{Lim-CVPRW-2017} + MMHCA (4 heads, r = $2$)                   & $42.05 / 0.9915$    \\ 
  \bottomrule
\end{tabular}
}
\end{center}
\vspace{-0.1cm}
  \caption{Ablation results with EDSR \cite{Lim-CVPRW-2017} on the NAMIC data set, for a scaling factor of $2\times$ and the T2w target modality. We consider different configurations for MHCA/MMHCA, varying the number of heads, the number of inputs, and the size of kernels.}
  \label{tab_namic_ablation}
\end{table}

\subsection{Results}
\vspace{-0.1cm}

Following~\cite{Feng-MICCAI-2021b,Lyu-TMI-2020,Zeng-CBM-2018,Zheng-BMCMI-2017,Zheng-ACCESS-2018}, we conduct experiments on the T2w target modality for the IXI and NAMIC data sets. The $^+$ sign after a method's name indicates the use of the geometric self-ensemble~\cite{Lim-CVPRW-2017}.

\noindent
\textbf{IXI.} We present the results obtained on the IXI data set for two upscaling factors, $2\times$ and $4\times$, in Table~\ref{tab_ixi}. The EDSR \cite{Lim-CVPRW-2017} model obtains a PSNR of $39.81$ and an SSIM of $0.9865$ for an upscaling factor of $2\times$. When we add the MHCA module to the single-contrast network, the performance increases to $40.11$ and $0.9871$ in terms of PSNR and SSIM, respectively. When we switch to the multimodal input, we observe performance improvements, regardless of the integrated attention module (MCSAM \cite{Feng-MICCAI-2021b,Niu-ECCV-2020},  MCBAM \cite{Woo-ECCV-2018} or MMHCA). This confirms our hypothesis that the information from the multi-contrast LR images is useful.
By adding MMHCA, the performance improves even further, exceeding the performance of both MCSAM and MCBAM. When EDSR is employed as underlying model, MMHCA$^+$ brings significant gains, generally outperforming the state-of-the-art methods \cite{Dong-TPAMI-2016,Feng-MICCAI-2021a,Georgescu-ACCESS-2020,Hui-CVPR-2018,Kim-CVPR-2016,Lim-CVPRW-2017,Shi-JBHI-2019,Zhang-CVPR-2021,Zhang-CVPR-2018,Zhao-TIP-2019}. 

In Figure~\ref{fig_2}, we illustrate qualitative results obtained by two baselines (bicubic and CNN+ESPC \cite{Georgescu-ACCESS-2020}) versus two enhanced versions of CNN+ESPC \cite{Georgescu-ACCESS-2020}, namely CNN+ESPC \cite{Georgescu-ACCESS-2020} + MHCA and CNN+ESPC \cite{Georgescu-ACCESS-2020} + MMHCA, for an upscaling factor of $4\times$. We observe that our CNN+ESPC \cite{Georgescu-ACCESS-2020} + MMHCA model obtains superior SR results compared with the baselines (bicubic, CNN+ESPC \cite{Georgescu-ACCESS-2020}), being able to recover structural details that are completely lost by the baselines. 

\noindent\textbf{NAMIC.} We show the results obtained on the NAMIC data set for two upscaling factors, $2\times$ and $4\times$, in Table~\ref{tab_namic}. The baseline CNN+ESPC \cite{Georgescu-ACCESS-2020} obtains a score of $33.92$ in terms of PSNR and $0.9509$ in terms of SSIM, for an upscaling factor of $4\times$. 
When we add MHCA and MMHCA, the scores improve by considerable margins (in terms of PSNR, the minimum improvement is $0.69$), regardless of the scale ($2\times$ or $4\times$) or the underlying model (EDSR~\cite{Lim-CVPRW-2017} or CNN+ESPC~\cite{Georgescu-ACCESS-2020}). 

\noindent
\textbf{Coltea-Lung-CT-100W.} We show the results obtained on the Coltea-Lung-CT-100W for two upscaling factors, $2\times$ and $4\times$, in Table~\ref{tab_ch}. 
Once again, we observe performance improvements brought by the use of multi-contrast LR inputs, the only attention that does not increase the baseline performance being MCBAM \cite{Woo-ECCV-2018}. Our MMHCA module exceeds the baseline and the other attention modules by significant margins (in terms of PSNR, the minimum improvement is $0.47$), regardless of the scale or the network. 


\noindent
\textbf{Quality assessment by physicians.}
In Table~\ref{tab_human}, we present the results of our subjective human evaluation study based on 100 cases, which are randomly selected from the IXI test set, for EDSR~\cite{Lim-CVPRW-2017}, with and without MMHCA, considering an upscaling factor of $4\times$. The quality evaluation study was completed by three physicians with expertise in radiology. The doctors had to choose between two images (randomly positioned on the left side or right side of the ground-truth HR image), without knowing which method produced each image. The HR images obtained after adding MMHCA were chosen in a proportion of $81.3\%$ against the images produced by the baseline EDSR~\cite{Lim-CVPRW-2017}. Upon disclosing the method producing each image, the doctors concluded that MMHCA helps to recover important details, \eg blood vessels, which are missed by the baseline EDSR.

\noindent
\textbf{Ablation study.}
In Table~\ref{tab_namic_ablation}, we present an ablation study on the NAMIC data set for a scaling factor of $2\times$. We observe that the best performance is obtained when the number of heads is equal to $3$, regardless of the number of input modalities (MHCA or MMHCA). We also notice that every configuration of our MMHCA module obtains better results than the concatenation of the features without any attention (multimodal input). 
To demonstrate the utility of the bottleneck principle, we test an ablated version of MHCA and MMHCA based on 3 heads, that does not reduce the input tensor through convolution, hence removing the deconv layer. This ablated version (3 heads, no deconv) attains lower performance compared to the complete version of MHCA and MMHCA based on 3 heads. This experiment supports our design based on the bottleneck principle through the conv and deconv operations.
To show that the diversity of the heads is important, we compare two modules with $4$ heads each and a reduction ratio of $r=2$, one where all kernel sizes are $1\times1$, and another where the kernels are of different sizes, from $1\times1$ to $7\times7$. The module based on diverse heads leads to significantly better results, indicating that using the same kernel size for all heads is suboptimal. In terms of the number of parameters, the module with 4 heads, $r=2$ and kernels of $1\times1$ is equivalent to the module with 1 head and $r=0.5$. 
The module with 1 head attains superior results, showing that simply adding more heads of the same kind is not useful. In summary, the empirical results show that our method obtains superior results not because of the increased capacity of the model, but due to the diversity of the heads.

\begin{figure}[!th]
  \centering
  \includegraphics[width=0.99\linewidth]{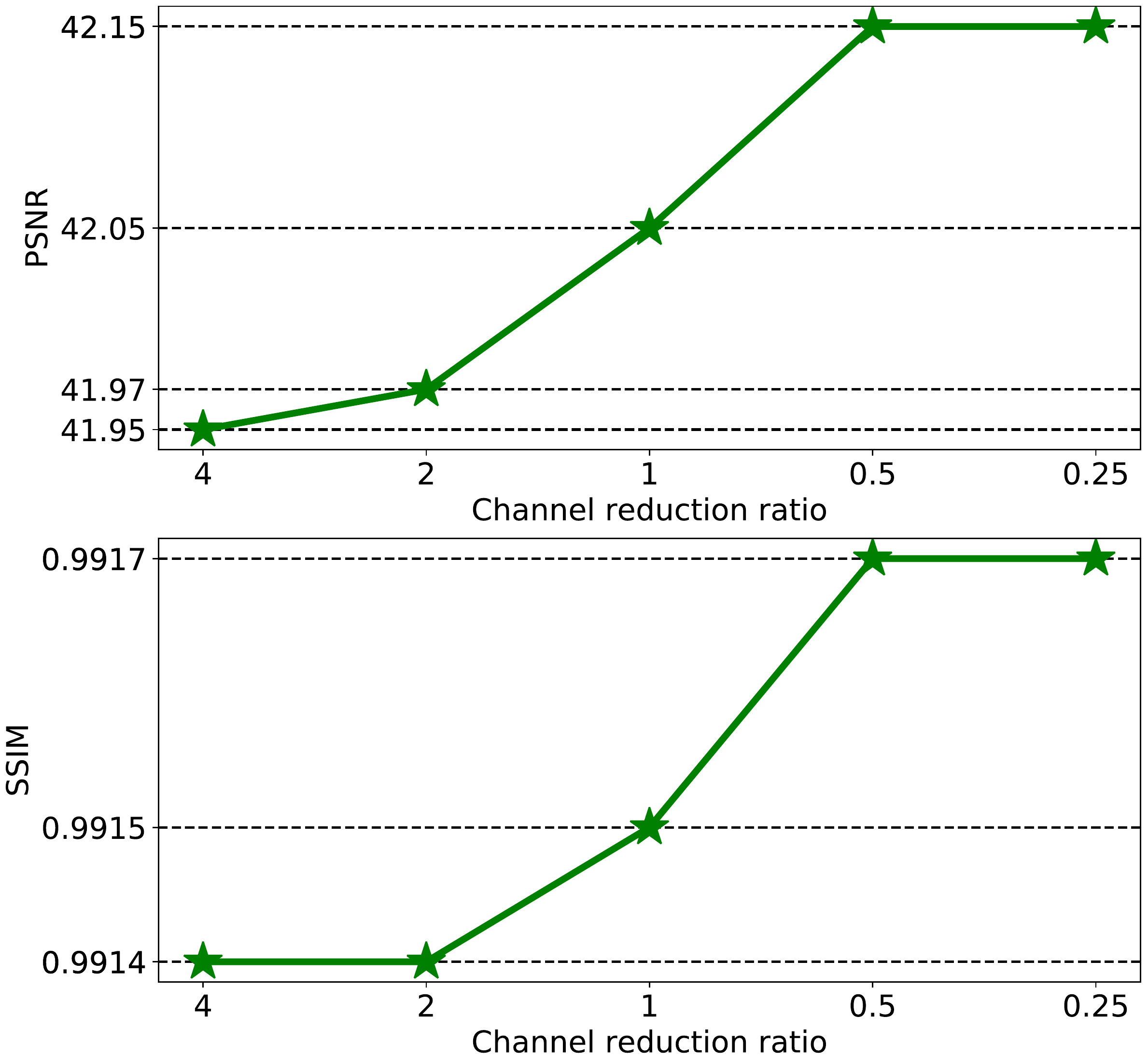}
  \caption{PSNR and SSIM scores of the EDSR \cite{Lim-CVPRW-2017} + MMHCA method, considering channel reduction rates in the set $\{4,2,1,0.5,0.25\}$. Results are reported on the NAMIC data set for an upscaling factor of $2\times$.}\label{fig_4}
  \vspace{-0.3cm}
\end{figure}

In Figure~\ref{fig_4}, we illustrate the impact of the channel reduction ratio on the MMHCA module based on 3 heads. We observe that the performance increases with the ratio until $r=0.5$. After this point, further increasing the ratio does not improve performance. 

\section{Conclusion}
\vspace{-0.1cm}

In this paper, we presented a novel multimodal multi-head convolutional attention (MMHCA) module, which performs joint spatial and channel attention. We integrated our module into two neural networks \cite{Georgescu-ACCESS-2020,Lim-CVPRW-2017} and conducted experiments on three multimodal medical image benchmarks: IXI, NAMIC and Coltea-Lung-CT-100W. We showed that our attention module yields higher gains compared with competing attention modules \cite{Feng-MICCAI-2021b,Niu-ECCV-2020,Woo-ECCV-2018}, being able to bring the performance of the underlying models above the state-of-the-art results \cite{Dong-TPAMI-2016,Feng-MICCAI-2021a,Georgescu-ACCESS-2020,Hui-CVPR-2018,Kim-CVPR-2016,Lim-CVPRW-2017,Shi-JBHI-2019,You-TMI-2019,Zeng-CBM-2018,Zhang-CVPR-2021,Zhang-CVPR-2018,Zhao-TIP-2019}.

In future work, we aim to integrate our module into further neural models and extend its applicability to natural images. We will also study the utility of the SR results in better solving other tasks, \eg medical image segmentation.

\section*{Acknowledgments}

The research leading to these results has received funding from the NO Grants 2014-2021, under project ELO-Hyp contract no. 24/2020. 

{\small
\bibliographystyle{ieee_fullname}
\bibliography{references}
}

\begin{figure*}[!th]
  \centering
    \includegraphics[width=0.98\textwidth]{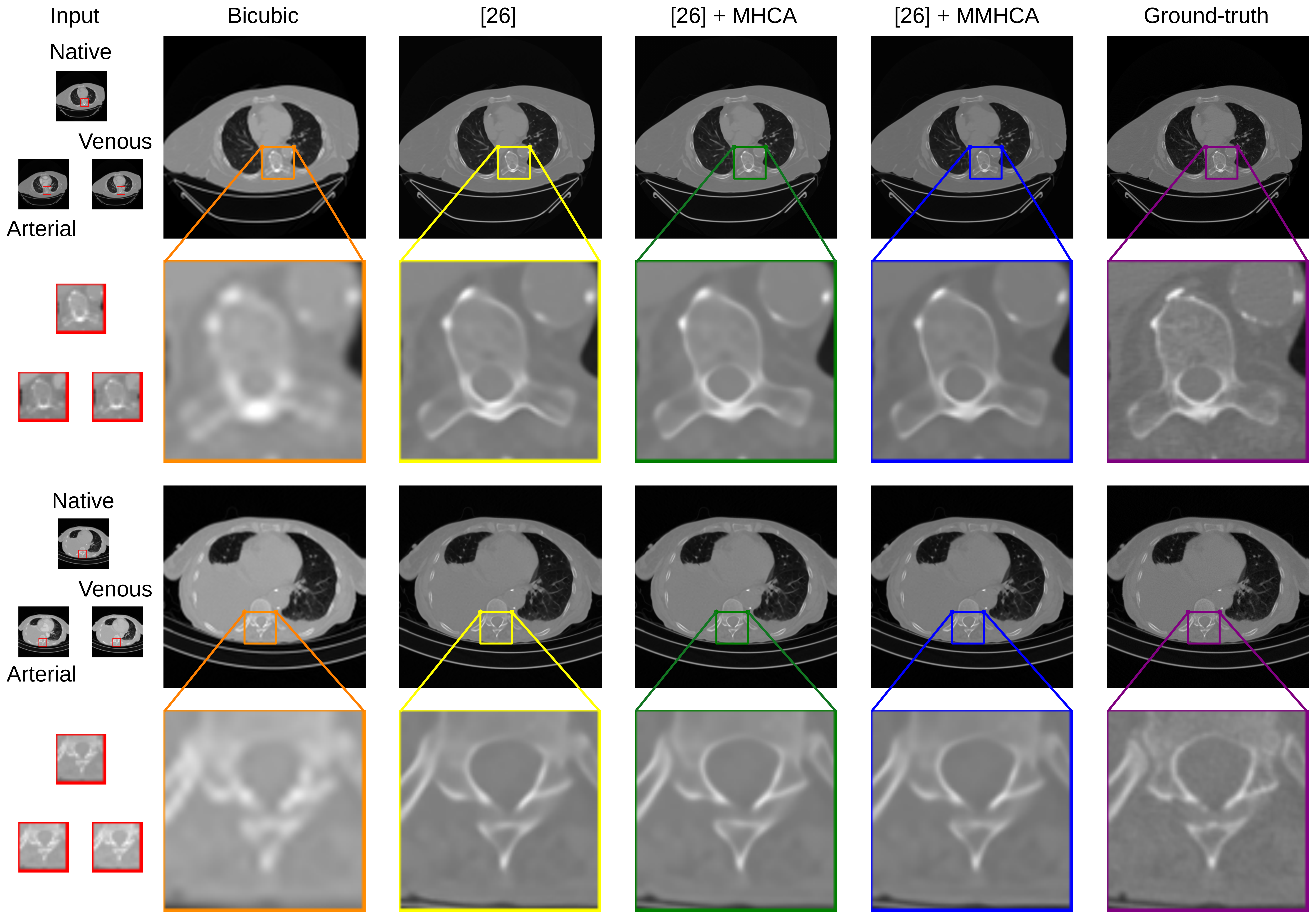}
    \caption{Examples of super-resolved CT images from the Coltea-Lung-CT-100W data set, for an upscaling factor of $4\times$. The HR images produced by two baselines (bicubic interpolation and EDSR \cite{Lim-CVPRW-2017}) are compared with the images given by two enhanced versions of EDSR \cite{Lim-CVPRW-2017}, one based on our single-contrast attention module (MHCA), and another based on our multimodal attention module (MMHCA).}
    \label{fig_3}
\end{figure*}

\begin{figure*}[!th]
  \centering
  \includegraphics[width=1.0\linewidth]{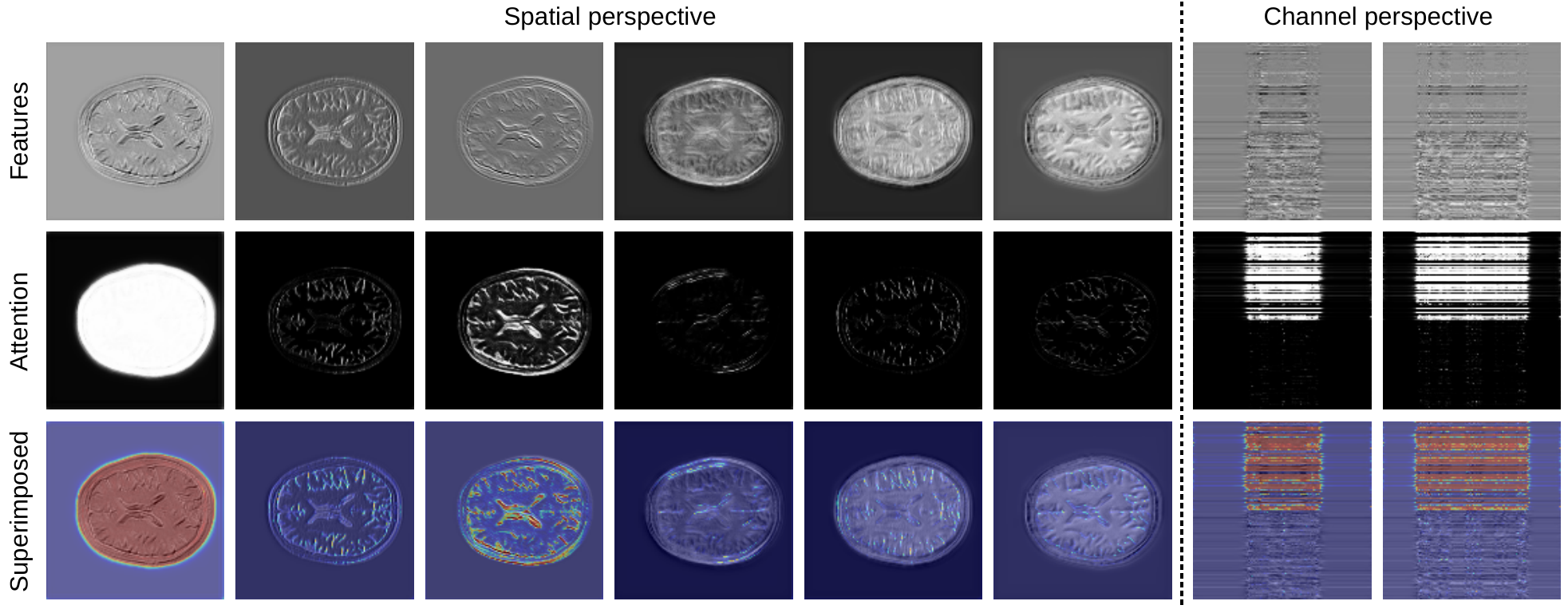}
  \caption{Views (top row) of a tensor computed for an example from NAMIC, which is given as input to MMHCA, and the corresponding attention maps (middle row) along the spatial (first six columns) and channel (last two columns) dimensions, showing that MMHCA performs joint channel and spatial attention. Views with superimposed attention maps are displayed on the bottom row.  Best viewed in color.} \label{fig_att}
\end{figure*}

\section{Supplementary}


\subsection{Additional Qualitative Results}

In Figure~\ref{fig_3}, we illustrate qualitative results obtained by two baselines (bicubic and EDSR \cite{Lim-CVPRW-2017}) versus two enhanced versions of EDSR \cite{Lim-CVPRW-2017}, namely EDSR \cite{Lim-CVPRW-2017} + MHCA and EDSR~\cite{Lim-CVPRW-2017} + MMHCA, for an upscaling factor of $4\times$. We observe that our EDSR \cite{Lim-CVPRW-2017} + MMHCA model is able to create sharper reconstructions and to improve the contrast levels.

\subsection{Attention Visualization}

In Figure~\ref{fig_att}, we show various perspectives along the spatial and channel dimensions of a tensor given as input to MMHCA. Looking at the attention corresponding to individual activation maps (first six columns), we observe that our module attends to salient contours and edges, or even full organs. Analyzing the attention along the channel axis (last two columns), we observe that, in this example, our module tends to mainly focus on the first LR input, naturally because the first LR input is the modality (contrast type) that corresponds to the HR output, containing the most relevant information to super-resolve the image. In contrast, the second modality is scarcely attended by our module. Overall, we observe that MMHCA performs both spatial and channel attention, confirming that our attention module works as intended.

\end{document}